\documentclass[twocolumn,pre,aps,superbib,tightenlines,floatfix,superscriptaddress,showpacs]{revtex4}
\usepackage[T1]{fontenc}
\usepackage[latin9]{inputenc}
\usepackage{amsmath}
\usepackage{graphicx}
\usepackage{amssymb}
\usepackage{hyperref}
\usepackage{color}

\begin{document}

\title{Complex dynamics and scale invariance of one-dimensional memristive networks}

\author{Yuriy V. Pershin}
\email{pershin@physics.sc.edu} \affiliation{Department of Physics and Astronomy and University of South Carolina Nanocenter, University of South Carolina, Columbia, South Carolina 29208, USA}
\author{Valeriy A. Slipko}
\email{slipko@univer.kharkov.ua} \affiliation{Department of Physics and Technology, V. N. Karazin
Kharkov National University, Kharkov 61077, Ukraine}
\author{Massimiliano Di Ventra}
\email{diventra@physics.ucsd.edu} \affiliation{Department of Physics, University of California, San Diego, California 92093-0319, USA}

\begin{abstract}
Memristive systems, namely resistive systems with memory, are attracting considerable attention due to their ubiquity in several phenomena and technological applications. Here, we show that even
the simplest one-dimensional network formed by the most common memristive elements with voltage threshold bears non-trivial physical properties. In particular, by taking into account the single element variability we find {\it i}) dynamical acceleration and slowing down of the total resistance in adiabatic processes, {\it ii}) dependence of the final state
on the history of the input signal with same initial conditions, {\it iii}) existence of switching avalanches in memristive ladders, and {\it iv}) independence of the dynamics voltage threshold with respect to the number of memristive elements in the network (scale invariance). An important criterion for this scale invariance is the presence of memristive systems with very small
threshold voltages in the ensemble. These results elucidate the role of memory in complex networks and are relevant to technological applications of these systems.
\end{abstract}

%Uncomment for PACS numbers title message
% Keywords required only for MST, PB, PMB, PM, JOA, JOB?
%\vspace{2pc}
%\noindent{\it Keywords}: Article preparation, IOP journals
% Uncomment for Submitted to journal title message
%\submitto{\JPA}
% Comment out if separate title page not required
\maketitle

\section{Introduction}

Resistors whose resistance depends on the past evolution of the system are quite common in both basic and applied science, and they have been known for at least several decades \cite{pershin11a}. Recently, they have attracted considerable interest in the context of memory applications -- where they are oftentimes referred to as memristive systems~\cite{chua76a} -- but their range of applicability spans various disciplines as diverse as non-traditional computing and biophysics \cite{pershin12a,Johnsen11a,pershin09b}. In addition
to their ubiquity, their theoretical description is quite simple: if a memristive system is subjected to a voltage $V(t)$, its resistance can be written as $R(x,V,t)$, namely it may depend on the voltage itself (which would simply make it a non-linear element), and, most importantly, it depends on some state
variable(s), $x$, which could be, e.g., the spin polarization \cite{pershin08a,wang09a} or the position of atomic defects \cite{yang08a}, or any other physical property of the system that gives it memory.

While most of the research so far has focused on the properties of these single elements -- both by identifying the
physical mechanisms for memory, and by devising innovative ways to employ them in practice -- very little is
known about their response when they are organized into networks, with the associated (and inevitable) element variability. In other words, the statistical properties of networks of memristive systems are largely unknown and, as we demonstrate below, are different than those of traditionally studied networks \cite{Albert02a,netw_book,motter12a}.

Addressing this issue has several immediate benefits. It is not at all obvious how a network whose elements have
memory of past dynamics evolves collectively in time, or whether it possesses fundamental (and universal)
characteristics that can be found in the natural world. In addition, a resistance with memory is just a particular case of
a general class of response functions that depend on state variables \cite{diventra09a} -- memcapacitive and meminductive \cite{diventra09a} systems are similarly defined. Moreover, networks with memory have been shown to be promising candidates for the solution of complex optimization problems (see, e.g., Ref. \onlinecite{pershin11d}), and their circuit applications are likely to involve a combination of several elements with memory. Finally, it is worth mentioning that the human (and animal) brain is - in its most basic description - simply a network with memory. This type of research may thus have consequences in neuroscience.
\begin{figure}[b]
 \begin{center}
    \includegraphics[width=7cm]{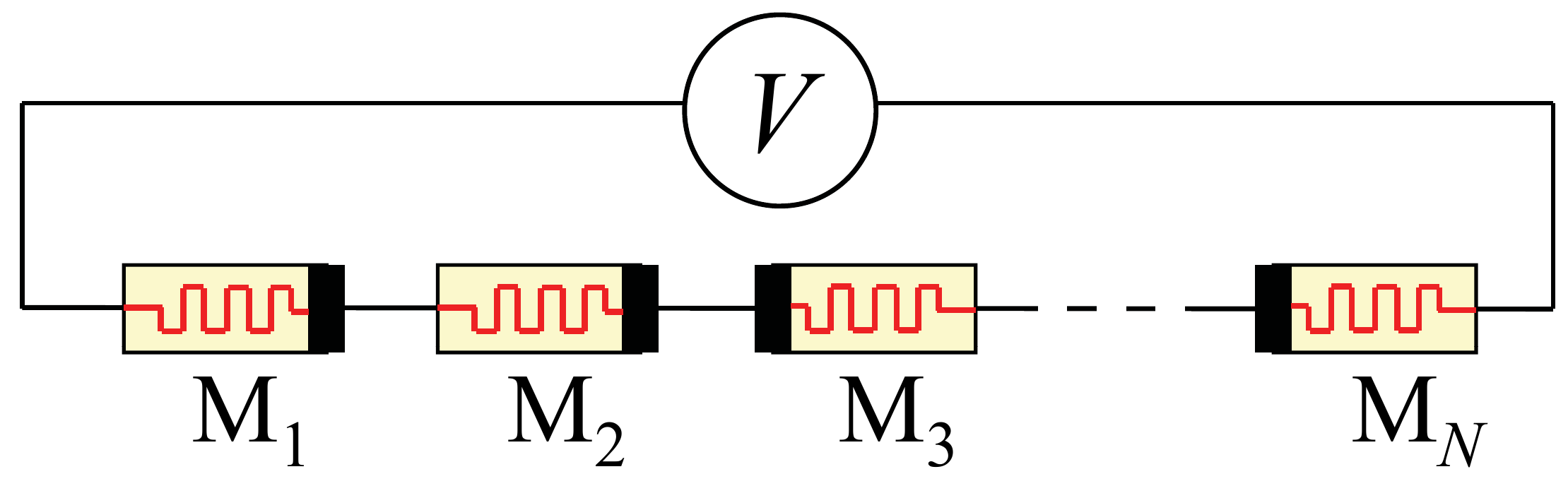}
\caption{\label{fig1} (color online). One-dimensional network of $N$ randomly oriented memristive systems M$_i$ connected to a voltage source $V$.}
\end{center}
\end{figure}

With these motivations in mind, in this paper we set to study, both numerically and theoretically, the statistical properties of the simplest network of memristive systems: a finite 1D network (Fig.~\ref{fig1}). In order to obtain its general features, we abstract as much as possible, and do not consider any particular
physical mechanism for memory. However, due to physical constraints, memristive systems are commonly found
with a threshold voltage, namely it takes a minimal voltage $V_{t}$ to change their resistance (see, e.g., Refs. \onlinecite{Jo09a,Schindler09b,strukov09a,pershin11a}). Therefore, in order to be as close as possible to experimental realizations, we focus on such type of elements. Moreover, it is generally observed that the memristance $R$ changes between two limiting values, $R_{on}$ and $R_{off}$. This experimental observation is also taken into account in our model. An extensive list of bipolar memristive devices with threshold can be found in the recent comprehensive review paper~\cite{pershin11a}. Specific examples of memristive systems with threshold that can be used to verify our predictions include different
 metal/oxide/metal memristive nanodevices~\cite{Jo09a,Schindler09b,yang08a}.

Probability distribution functions $f_{V_t}(V)$, $f_{R}(R)$, $f_{R_{on}}(R)$, $f_{R_{off}}(R)$ satisfying the normalization conditions
\begin{equation}
\int_0^\infty f_{i}(y)\textnormal{d}y=1 \label{eq_normal}
\end{equation}
where $i=V_t$, $R$, $R_{on}$ or $R_{off}$ are used to describe threshold voltages, memristances and limiting values of memristances in the network, respectively. To understand the meaning of these distribution functions, we note that, e.g., $f_{V_t}(V)\textnormal{d}V$ is the probability to find a memristive device in the network with $V_t$ in the interval between $V$ and $V+\textnormal{d}V$.
All other distribution functions are introduced in a similar way.
Moreover, we suppose that systems with required $f_{V_t}(V)$, $f_{R_{on}}(R)$ and $f_{R_{off}}(R)$ can be fabricated "on demand". (For example, one can extract a subset of devices from an ensemble with Gaussian distribution to form a different distribution.) Even in an ensemble of devices fabricated with specific values of $V_t$, $R_{on}$ or $R_{off}$, an inevitable fluctuation of $V_t$, $R_{on}$ and $R_{off}$ around their average values can also be described by the above mentioned distribution functions.
Finally, we note that the memristance distribution function $f_{R}(R)$ is a time-dependent function, unlike  $f_{V_t}(V)$, $f_{R_{on}}(R)$ and $f_{R_{off}}(R)$ distribution functions. The distribution function $f_{R}(R)$ can be easily pre-set to the required form by applying appropriate pulse sequences to individual memristive devices~\cite{pershin09d}. Below, we give several examples of the evolution of $f_{R}(R)$.

\begin{figure}[tb]
 \begin{center}
    \includegraphics[width=6.5cm]{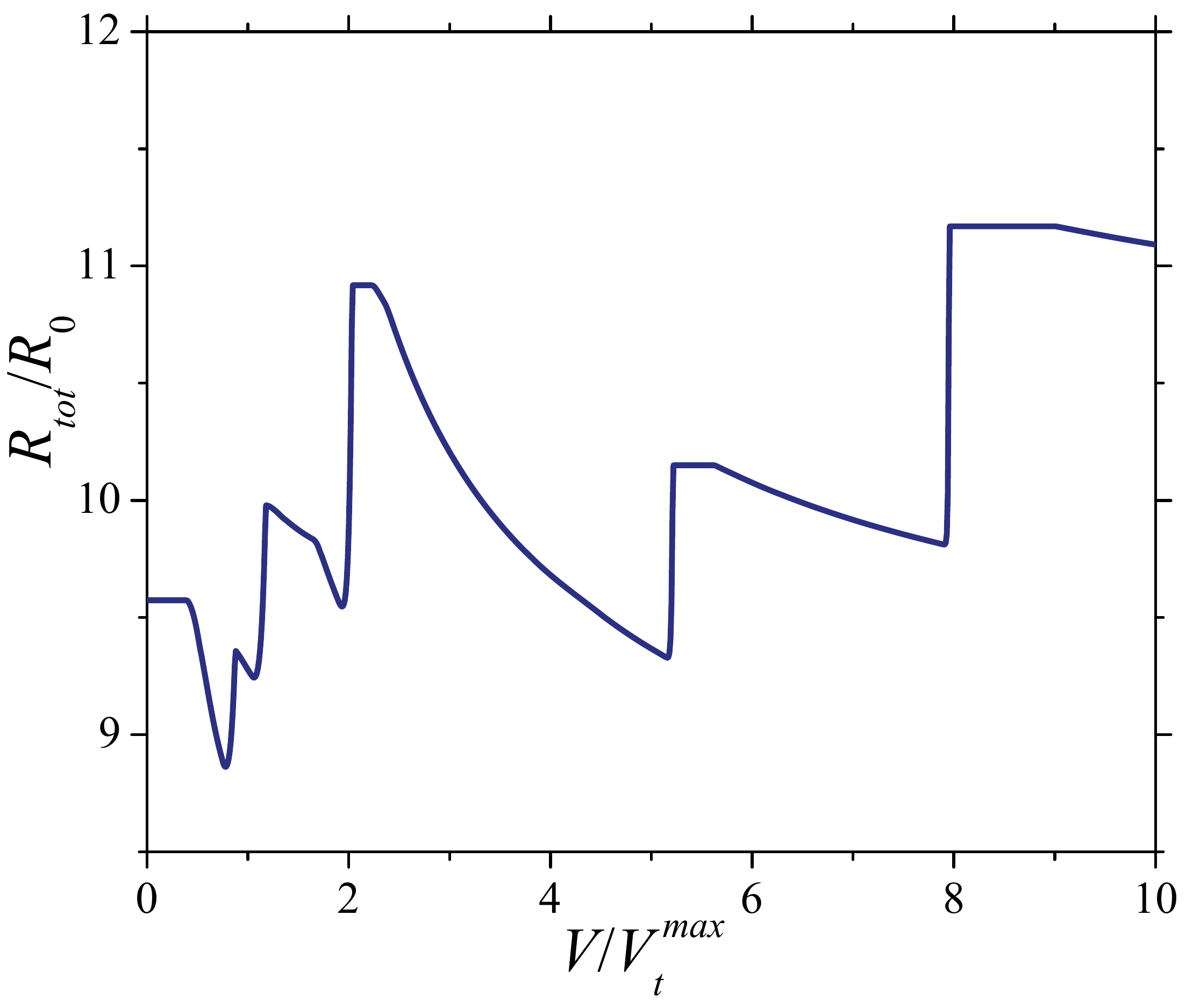}
\caption{\label{fig4} (color online). Total memristance $R_{tot}$ in a network of 10 randomly oriented devices as a function of a slowly ramped voltage
$V$. In this simulation, we have used fixed values of $R_{on}=0.05R_0$ and $R_{off}=1.95R_0$, and $\beta=2R_0/\left(V_t^{max}\cdot \textnormal{s} \right)$. The threshold voltages and initial memristances of memristive systems have been randomly selected in the interval [0,$V_t^{max}$] and [$R_{on}$,$R_{off}$], respectively.}
\end{center}
\end{figure}

In what follows, we will show that despite their apparent simplicity, memristive networks indeed show a quite complex dynamical behavior, and some general scale-invariant properties that can be readily verified experimentally. This paper is organized as follows. In Sec. \ref{sec2} we consider adiabatic processes and introduce two types of memristive dynamics. Sec. \ref{sec3} studies the dependence of the final state on the history of applied voltage. Scale-invariance properties of memristive networks are demonstrated in Sec. \ref{sec4}. In Sec. \ref{sec5} we consider a memristive ladder and demonstrate the existence of avalanches in its switching dynamics. Finally, in Sec. \ref{sec6} we report our conclusions.

\section{Acceleration and slowing-down of the resistance switching in adiabatic processes} \label{sec2}

Let us first discuss the switching dynamics in a 1D chain of memristive systems. For the sake of definiteness let us consider a simplified version of a generic threshold voltage model of memristive system~\cite{pershin09b}
\begin{eqnarray}
I&=&x^{-1}V_M, \label{eq3} \\ \dot x&=&\beta \left( V_M-0.5\left[ |V_M+V_{t}|-|V_M-V_{t}| \right]\right)
\nonumber\\ & &\times \left[\theta\left( V_M\right) \theta\left(
R_{off}-x\right)+ \theta\left(- V_M\right) \theta\left(
x-R_{on}\right)\right] \quad \label{eq4},
\end{eqnarray}
where $I$ and $V_M$ are the current through and the voltage across the system, respectively,
$x$ is the internal state variable playing the role of memristance, $x \equiv R$,
$\theta(\cdot)$ is the step function, $\beta$ is a positive constant
characterizing the rate of memristance change when
$|V_M|> V_{t}$,
$V_t$ is the threshold voltage, and  $R_{on}$ and $R_{off}$ are limiting
values of the memristance $R$. In Eq. (\ref{eq4}), the role of $\theta$-functions
is to confine the memristance change to the interval between
$R_{on}$ and $R_{off}$. In our numerical simulations, the value of $x= R$  is monitored at each time step
and in the situations when $x<R_{on}$ or $x>R_{off}$, it is set equal to
$R_{on}$ or $R_{off}$, respectively. Importantly, the model given by Eqs. (\ref{eq3}) and (\ref{eq4}) describes a system whose memristance can only change when the applied voltage magnitude exceeds the threshold voltage. In addition, according to the sign of the applied voltage
its resistance may increase or decrease, thus representing an asymmetric system, as indicated by the black thick line in the memristor symbols in Fig.~\ref{fig1}.

Fig. \ref{fig4} presents the total memristance of a network of such elements as a function of a slowly ramped voltage for a particular network realization. The initial resistances and the threshold voltages have been chosen randomly within uniform distributions. The polarity of the memristive elements in the network has also been randomly selected. The voltage is then increased following an {\it adiabatic process}: its increase rate is so slow that at any moment of time the network is in its equilibrium state. However, even in this simple adiabatic switch-on, one can easily notice two distinctive types of switching behavior. An abrupt -- accelerated --  switching, with well defined steps (see Fig. \ref{fig4}), occurs when the memristance $R_j$ of the $j$-th memristor in the network increases at the given voltage polarity. In fact, as soon as the voltage drop across such element exceeds its threshold voltage, the increase in memristance increases the voltage drop across it, thus accelerating the switching. Instead, for memristive systems connected with opposite polarity, a decrease in memristance decreases the voltage drop across the system, thus decelerating the switching.
This effect is clearly seen in long switching tails.

The dependence of the switching rate on the device polarity in 2D networks of ideal memristors~\cite{chua71a} was previously reported in Ref. \cite{Oskoee11a}. This previous result, however, does not take into account the threshold-type dynamics of realistic memristive devices~\cite{pershin11a}. Consequently, in adiabatic experiments, instead of sharp steps and long tails predicted by us, this previous work suggests unrealistic switching of memristors into their limiting states already at very small applied voltages and the absence of any subsequent evolution of $R_{tot}$ as shown in Fig. \ref{fig4}.

\begin{figure}[tb]
 \begin{center}
    \includegraphics[width=6.5cm]{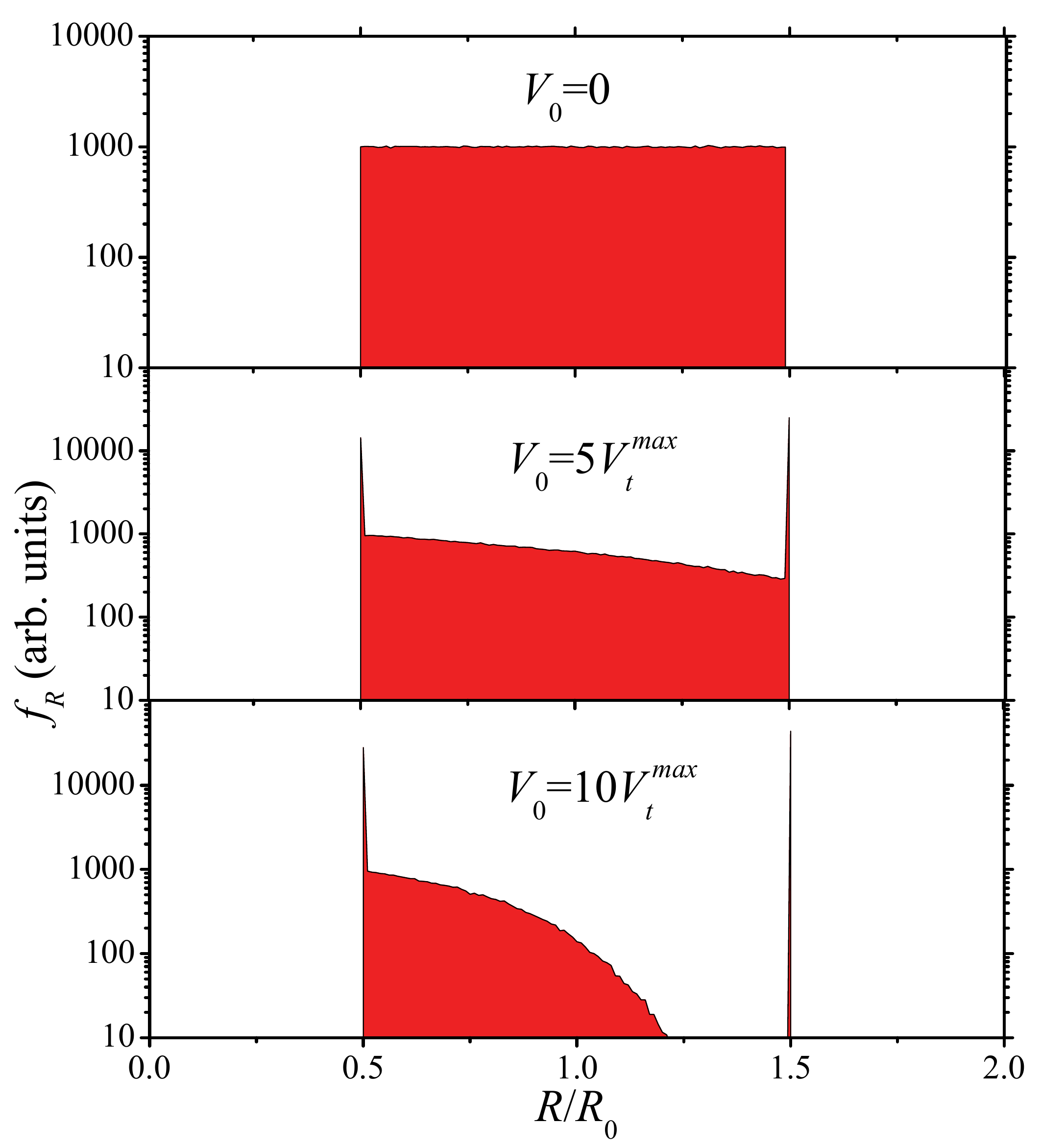}
\caption{\label{fig5} (color online). Evolution of memristance distribution function $f_R$ in a network of 10 randomly oriented memristive systems subjected to an adiabatically switched voltage $V_0$. In this simulation, we have used fixed values of $R_{on}=0.5R_0$ and $R_{off}=1.5R_0$, $\beta=2R_0/\left(V_t^{max}\cdot \textnormal{s} \right)$, the threshold voltages and initial memristances of memristive systems have been randomly selected in the interval [0,$V_t^{max}$] and [$R_{on}$,$R_{off}$], respectively. The memristance distribution function $f_R$ has been obtained by averaging over $10^4$ random realizations of the network.}
\end{center}
\end{figure}

\begin{figure}[tb]
 \begin{center}
    \includegraphics[width=6.5cm]{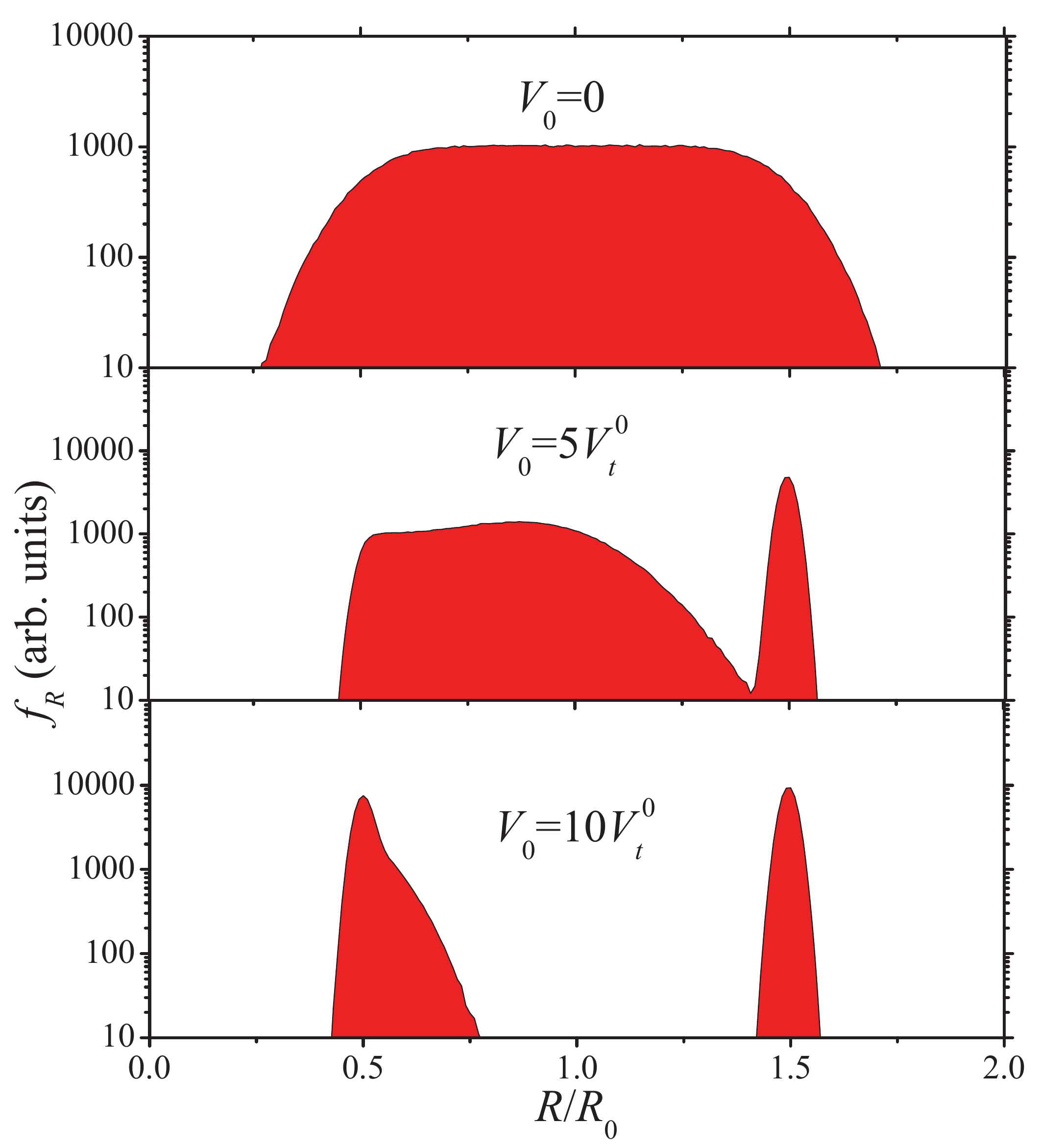}
\caption{\label{fig5a} (color online). Evolution of memristance distribution function $f_R$ in a network of 10 randomly oriented memristive systems subjected to an adiabatically switched voltage $V_0$. In this simulation, we have used normally distributed $R_{on}$, $R_{off}$, and $V_t$ with means $0.5R_0$, $1.5R_0$ and $V_t^{0}$ and standard deviations $0.02R_0$, $0.02R_0$ and $0.05V_t^{0}$, respectively, $\beta=2R_0/\left(V_t^{0}\cdot \textnormal{s} \right)$. The memristance distribution function $f_R$ has been obtained by averaging over $10^5$ random realizations of the network.}
\end{center}
\end{figure}

The accelerated/decelerated switching behavior described above reflects also on the evolution of the resistance distribution function, $f_R(x)$, of the network. This is shown in Fig. \ref{fig5} at several moments of time when the applied voltage $V_0$ adiabatically ramps up from 0 to 10$V_t^{max}$. In fact, the memristance distribution function dynamics (Fig. \ref{fig5}) indicates a faster switching of memristive systems with initially higher memristances (thus with voltage drops exceeding their threshold voltages).
Moreover, while the switching into the "off" state is always complete (since it is accelerated), memristive systems connected with opposite polarity may only partially switch toward the "on" state because of the decelerating switching effect discussed above. Higher values of applied voltage would further promote their switching as well as initiate switching of memristive systems with higher $V_t$ and lower initial resistances, requiring higher voltage drops to induce their dynamics.

Fig. \ref{fig5a} presents the evolution of the resistance distribution function in the case of normally distributed threshold voltages and limiting values of memristances. The overall behavior of $f_R(x)$ is very similar to that shown in Fig. \ref{fig5} for the case of uniformly distributed device parameters. Specifically, in both cases, memristive devices switch into the limiting $R_{off}$ states faster than into $R_{on}$ ones. Moreover, the dynamics of total memristance, $R_{tot}(t)$ in the  network with normally distributed device characteristics has exactly the same features as
$R_{tot}(t)$ depicted in Fig. \ref{fig4} (acceleration and slowing-down of the switching ).

\begin{figure}[tb]
 \begin{center}
    \includegraphics[width=7.5cm]{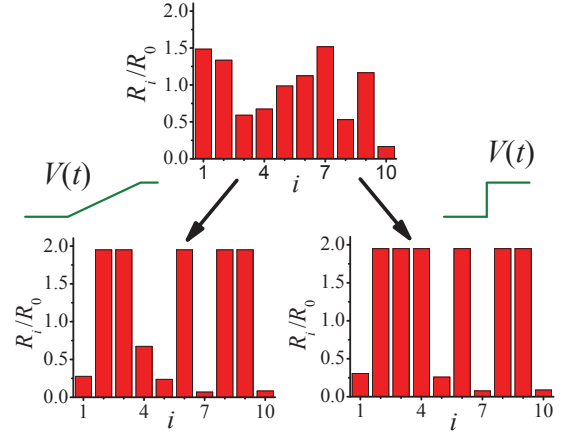}
\caption{\label{fig6} (color online). Initial (top) and final (bottom) memristances in a network of 10 memristive elements (denoted by the index $i$) subjected to a slow ramp-up and a step-like applied voltage so that in both cases $V(t=t_{final})=10V_t^{max}$ (the respective voltage shapes are indicated in the middle). In this simulation, we have used fixed values of $R_{on}=0.05R_0$ and $R_{off}=1.95R_0$, $\beta=2R_0/\left(V_t^{max}\cdot \textnormal{s} \right)$, the threshold voltages and initial memristances have been randomly selected in the interval [0,$V_t^{max}$] and [0.05$R_0$,1.95$R_0$], respectively, and the polarity of the memristive systems in the network has also been randomly selected. However, both simulations have been performed starting with the same initial randomly selected network realization.}
\end{center}
\end{figure}

\section{Different final states with same initial condition} \label{sec3}

Another consequence of memory is that the memristance at the final moment of time is determined both by the initial conditions {\it and} by the actual shape of the applied signal, so that if the initial conditions are the same, the final memristance could be different according to
the way the input signal is applied. This effect is even more pronounced in memristive networks, where all elements are {\it collectively} coupled
by current conservation. As an example we compare the evolution of memristances in a network subject to both a slowly ramped-up voltage (from 0 to 10$V_t^{max}$) and a step voltage of 10$V_t^{max}$ applied at $t=0$. The initial conditions (initial parameters of memristive systems in the network) are chosen to be the same. Fig. \ref{fig6} shows results of our simulations for a particular network realization of ten elements. One can clearly notice that in this particular situation the main difference of final states is associated with the fourth memristor from the left. While in the case of slowly ramped voltage $R_4(t=t_{final})=R_4(t=0)$, the sudden voltage pulse switches the fourth memristive element  into the "off" state $R_{off}$. According to our observations, such a significant difference occurs in $\sim 8\%$ of network realizations at the selected simulation conditions.

In order to understand the signal-dependent switching in the memristive networks,
let us consider the simplest nontrivial case of a network of two memristive
elements, $M_1$ and $M_2$,
connected in  series (see Fig. \ref{fig1}) .
According to Eqs. (\ref{eq3}), (\ref{eq4}), such a network, subjected to a non-negative voltage $V\geq 0$, is described by the following equations
\begin{eqnarray}
\frac{dR_1}{dt}=\begin{cases}
\pm\beta\left( \frac{R_1}{R_1+R_2}V-V_{t1} \right),
\mbox{when }\frac{R_1}{R_1+R_2}V> V_{t1}, \\
\mbox{and } R_1<R_{off} \mbox{ for sign "+",}\\
\mbox{or } R_1>R_{on}  \mbox{ for sign "-";}\\
0, \mbox{in all other cases}; \\
\end{cases} \label{i1} \\
\frac{dR_2}{dt}=\begin{cases}
\pm\beta\left( \frac{R_2}{R_1+R_2}V-V_{t2} \right),
\mbox{when }\frac{R_2}{R_1+R_2}V> V_{t2}, \\
\mbox{and } R_2<R_{off} \mbox{ for sign "+",}\\
\mbox{or } R_2>R_{on}  \mbox{ for sign "-";}\\
0, \mbox{in all other cases}, \\
\end{cases} \label{i2}
\end{eqnarray}
where $R_1$ and $R_2$, and $V_{t1}$ and $V_{t2}$  are the memristances and the threshold voltages  of the first and second memristive systems, respectively. The signs $"\pm"$ in the RHS
of Eqs. (\ref{i1}), (\ref{i2}) take into account different possible polarities
of the memristive systems in the network. Here we suppose that the parameter $\beta$
determining the switching rate is positive and of the same magnitude for both
memristive systems.  Eqs. (\ref{i1}), (\ref{i2}) determine a system of  nonlinear differential equations, which must be supplied with the initial conditions for memristances
$R_1$ and $R_2$ at $t=0$.

Let us consider a situation when a constant voltage $V$ is applied at $t=0$. In this case, Eqs. (\ref{i1}), (\ref{i2}) can be solved analytically. However, due to rather complex functional dependencies in the RHS of these equations, there are a lot of cases
depending on the mutual relations between the initial memristances, threshold voltages, and applied voltage $V$. Below, we investigate several possible scenarios for signal-dependent switching.

According to Eqs. (\ref{i1}), (\ref{i2}), any changes
in memristances occur when the voltage drop across at least one memristive system exceeds its threshold voltage. Let us consider then the case when the voltage drop $V_1=R_{1}V/(R_1+R_2)$ across $M_1$ exceeds $V_{t1}$, but the voltage drop $V_2=R_{2}V/(R_1+R_2)$ across $M_2$ is smaller then  $V_{t2}$.
We also suppose that $R_{on}<R_1<R_{off}$.  In this case (see Eq. (\ref{i2})), the memristance $R_2$ does not
change, $R_2=\textnormal{const}$. Eq.  (\ref{i1}) can be integrated resulting in the following algebraic equation for $R_1(t)$:
\begin{eqnarray}
R_1+\frac{V R_2}{V-V_{t1}}\ln\left[
\frac{(V-V_{t1})R_1-V_{t1}R_2}{(V-V_{t1})R_1(0)-V_{t1}R_2} \right]=\nonumber\\
R_1(0)\pm\beta(V-V_{t1}) t.
\label{i3}
\end{eqnarray}
It should be noted that Eq. (\ref{i3}) is valid while the following inequalities
are satisfied:
\begin{eqnarray}
R_{on}<R_1<R_{off},\label{i3.1}\\
V_1=\frac{R_{1}}{(R_1+R_2)}V>V_{t1},\label{i3.2}\\
V_2=\frac{R_{2}}{(R_1+R_2)}V<V_{t2}.\label{i3.3}
\end{eqnarray}
If we have a "positive" polarity with respect to applied external voltage $V$ (sign "$+$" before $\beta$ in the RHS of Eq. (\ref{i1})), then as it follows from Eq.  (\ref{i3})
or directly from Eq.  (\ref{i1}), the memristance $R_1$ increases with time. In
this case the voltage drop $V_1$ increases, but $V_2$ decreases, because
the total voltage drop across both memristors is constant, $V_1+V_2=V$.
Thus the memristance $R_1$ increases with an acceleration (as $V_1$ increases).
When $R_1(t)$ reaches $R_{off}$ (the corresponding moment of time $t$
can be immediately calculated
from Eq.(\ref{i3})), the memristance $R_1$ stops increasing
(according to Eq. (\ref{i1})) and stays constant.

In the case of a "negative" polarity
(sign "$-$" before $\beta$ in the RHS of Eq. (\ref{i1})), it follows from Eq.  (\ref{i3})
(or directly from Eq.  (\ref{i1})) that the memristance $R_1$ decreases with time. In
this case the voltage drop $V_2$ increases, and $V_1$ decreases. This evolution, as
determined by
 Eq.(\ref{i3})
stops when memristance  $R_1$ reaches  the lowest possible value $R_{on}$,
or inequality (\ref{i3.2}) does not hold, depending on what happens first.
If inequality (\ref{i3.3}) becomes invalid first, while other inequalities (\ref{i3.1}),
(\ref{i3.2}) are still valid, we encounter the situation when both memristances
change (if, of course, $R_{on}<R_2<R_{off}$). Let us now consider this case when both memristances are changed simultaneously. This is possible only when the following inequalities are valid:
\begin{eqnarray}
R_{on}<R_{1,2}<R_{off},\label{i4}\\
V_1=\frac{R_{1}}{(R_1+R_2)}V>V_{t1},\label{i5}\\
V_2=\frac{R_{2}}{(R_1+R_2)}V>V_{t2}.\label{i6}
\end{eqnarray}

We limit ourselves to the cases when the memristive
elements, $M_1$ and $M_2$, are
connected
with the same polarity. Then by taking the sum of Eqs. (\ref{i1}) and  (\ref{i2}),
and integrating it with respect to time we find for the total memristance
\begin{eqnarray}
R_1+R_2=\pm\beta(V-V_{t1}-V_{t2})t+R_1(0)+R_2(0).
\label{i7}
\end{eqnarray}
By substituting this relation
into Eq. (\ref{i1}) we get a linear differential equation which can be  easily integrated. As a result we find the following expressions which determine
time dependence of memristances for both memristors:\
\begin{eqnarray}
R_1(t)=P(R_{1}+R_{2})^{\frac{V}{V-V_{t1}-V_{t2}}}+\frac{V_{t1}}{V_{t1}+V_{t2}}(R_{1}+R_{2}),~~
\label{i8}
\end{eqnarray}

\begin{eqnarray}
R_2(t)=-P(R_{1}+R_{2})^{\frac{V}{V-V_{t1}-V_{t2}}}+\frac{V_{t2}}{V_{t1}+V_{t2}}(R_{1}+R_{2}),~
\label{i9}
\end{eqnarray}
 where parameter $P$ is determined by the initial conditions
\begin{eqnarray}
P=\frac{V_{t2}R_{1}(0)-V_{t1}R_{2}(0)}{V_{t1}+V_{t2}}
\left( R_1(0)+R_2(0) \right)^{-\frac{V}{V-V_{t1}-V_{t2}}}.~~
\label{i10}
\end{eqnarray}

From Eqs. (\ref{i7})-(\ref{i10}) it follows that the voltage drop $V_1$ increases with time
for  the "positive" polarity and when
$V_{t2}R_{1}(0)>V_{t1}R_{2}(0),$ while voltage drop $V_2$ decreases. Thus
it leads to accelerating of the switching of memristor $M_1$ and to decelerating of the switching of memristor $M_{2}$.

Having discussed the switching dynamics due to a constant applied voltage, let us compare the final states of memristive systems for two different bias application protocols. Specifically, assuming the same initial states of the memristive systems and the same final voltage magnitude, we consider the situations of the step and slowly ramped voltages. Our consideration is based on the following model parameters: $R_1(0)=R_2(0)=0.2R_0$,
$R_{on}=0.05R_0$, $R_{off}=1.95R_0$, $V_{t1}=V/3$,  $V_{t2}=1.1V/3$, and
the signs $"+"$ in the RHS
of Eqs. (\ref{i1}), (\ref{i2}).

For slowly ramped applied voltage, $M_1$ is in the regime of accelerated switching (see also discussion after Eq. (\ref{i3.3})). In addition, one can easily notice that
 the voltage drop across $M_2$ never exceeds $V_{t2}$.
 Consequently, the final memristances are
$R_1(t=t_{final})=R_{off}=1.95R_0$ and $R_2(t=t_{final})=R_2(t=0)=0.2R_0$.

In the case of the step-like applied voltage $V$, with our specific model parameters the
inequalities (\ref{i4})-(\ref{i6})
are satisfied at the initial moment of time. As a result, both memristances are changing according to Eqs. (\ref{i7})-(\ref{i10}).
Because of the positivity of parameter $P$, the voltage $V_1$ increases in time, but the voltage  $V_2$ decreases
(as it follows from Eqs. (\ref{i7})-(\ref{i9})). At the
same time, both memristances $R_1$ and $R_2$ increase  according to  Eqs. (\ref{i1}), (\ref{i2}). At a certain moment of time the voltage drop $V_2$ across memristor $M_2$ becomes equal to the threshold voltage $V_{t2}$, and further change
of $M_2$ becomes impossible. Further evolution of memristance $R_1$ is determined by Eq.  (\ref{i3}) and lead to
$R_1(t=t_{final})=R_{off}=1.95R_0$, while the memristance $R_2$ does not change, $R_2(t=t_{final})=0.33R_0$. Clearly, the final state of $M_2$ in this case is different than that for the case of the slowly ramped voltage.

We would like to emphasize that the different final states are obtained as a result of the {\it collective dynamics} of many memristive systems in the network. Such a behavior cannot be observed considering a single memristive device described by Eqs. (\ref{eq3}), (\ref{eq4}) subjected to similar voltage patterns. The interaction among devices in the network is provided by the current that, at each moment of time, is determined by  states of {\it all} devices in the network.

Our observation of different final states can find useful applications in electronics, albeit in a modified form. For example, the sensitivity of the final states to the applied pulse profile can be used in signal recognition. We would also like to note that the response of single memristive devices with more complex internal degrees of freedom can be richer than that predicted by Eqs. (\ref{eq3}), (\ref{eq4}). Even with such single memory devices, one can expect the dependence of the final states on the voltage application protocol. Examples of such memory devices can be found in Refs. \cite{Driscoll10b,pershin10a}.

\section{Scale invariance of dynamics threshold voltage} \label{sec4}

In this Section, we ask the question of the probability of dynamics in the memristive network.  Clearly, if a very low voltage $V$ is applied to the network
then the probability of dynamics is close to zero as the voltage across each memristive system is likely below its threshold voltage. If a high voltage is applied, the probability of dynamics is close to 1 as it is highly probable that the voltage across at least one memristive system exceeds its threshold. To quantify this effect, let us find a {\it dynamics voltage threshold} $V^*$, which is the magnitude of the applied voltage when, with a probability of 1/2, the state of at least a single memristive system will change. Formally,
it can be found from
\begin{equation}
p_{d}(V^*)=1/2, \label{eq1}
\end{equation}
where $p_d$ is a probability of dynamics in the network. Alternatively, Eq. (\ref{eq1}) can be rewritten as $1-p_{d}(V^*)=1/2$, where the left-hand-side is the probability of no dynamics. The latter occurs when the voltage across each memristive system in the network, $V_i$, does not exceed its threshold voltage, $V_{t,i}$. Consequently, the probability of no dynamics can be written as a product of single element probabilities
\begin{equation}
p(V_1\leq V_{t,1})p(V_2\leq V_{t,2})...p(V_N\leq V_{t,N})=1/2, \label{eq33}
\end{equation}
where $V_i=R_iV^*/R_{tot}$ and $R_{tot}=\sum _{i=1}^N R_i$. Noticing that the probabilities in Eq. (\ref{eq33}) can be expressed through the threshold voltage distribution function $f_{V_t}$ as
\begin{equation}
p(V_i\leq V_{tr,i})=\int\limits_{V_i}^{\infty}f_{V_t}(y)\textnormal{d}y, \label{eq44}
\end{equation}
we finally arrive at the following equation
\begin{equation}
\int\limits_{V_1}^{\infty}f_{V_t}(y_1)\textnormal{d}y_1\int\limits_{V_2}^{\infty}f_{V_t}(y_2)\textnormal{d}y_2...
\int\limits_{V_N}^{\infty}f_{V_t}(y_N)\textnormal{d}y_N=1/2  \label{eq5}
\end{equation}
determining the dynamics threshold voltage $V^*$.

\subsection{General case, narrow distribution of initial memristances}

Assuming that initial memristances are narrow distributed around a certain value $\bar{R}$, we use $V_i=V^*/N$ in Eq. (\ref{eq5}) to derive
\begin{equation}
\left[\int\limits_{V^*/N}^{\infty}f_{V_t}(x)\textnormal{d}x\right]^N=1/2.
\label{eq9}
\end{equation}
Let us then find the dynamics threshold voltage $V^*$ in the limit of large $N$.
By using the normalization condition, and by taking the logarithm of both sides
of Eq. (\ref{eq9}) we find, for arbitrary $N$, the
following relation
\begin{equation}
\ln\left[1-\int\limits_{0}^{V^*/N}f_{V_t}(x)\textnormal{d}x\right]=-\frac{\ln 2}{N}. \label{eq10}
\end{equation}
From Eq. (\ref{eq10}) we conclude that for large $N$ the integral in Eq.
(\ref{eq10}) tends to zero.  Because of the positivity of the distribution function
$f_{V_t}(x)$ this means that the integration domain must tend to zero as $N$
tends to infinity. Thus we see that $\lim_{N\rightarrow\infty} V^*/N =0$.
This means that we can use the asymptotic behavior of the distribution  function
to find the dynamics threshold voltage. Let us then consider a general situation when
\begin{equation}
f_{V_t}(x)=f_0 x^\gamma,~~~x\rightarrow0, \label{eq11}
\end{equation}
with some positive parameters $f_0$ and $\gamma\geq 0$.
In this case the integration in Eq. (\ref{eq10}) gives the final result
\begin{equation}
V^*=\left[\frac{(1+\gamma)\ln 2}{f_0}\right]^{1/(1+\gamma)}N^{\frac{\gamma}{1+\gamma}}, \;\;\;\;\;\;
N\rightarrow \infty. \label{eq12}
\end{equation}

When there is a finite probability of arbitrarily small thresholds
($\gamma=0$ and $f_0\neq 0$), we obtain a scale invariant expression
\begin{equation}
V^*=\frac{\ln 2}{f_{V_t}(0)}, \;\;\;\;\;\; N\rightarrow \infty. \label{eq13}
\end{equation}
This is the main result of Section \ref{sec4}.

\subsection{Exponential distribution of threshold voltages}

Note that the condition $R_i=\bar R$ is not always necessarily required to obtain Eq. (\ref{eq13}). To see this,
let us consider, e.g., an exponential distribution of threshold voltages, $f_{V_t}(y)=\textnormal{exp}\left( -y/V_0 \right)/V_0$, where $V_0$ is a characteristic threshold voltage of memristive systems. It follows from Eq. (\ref{eq5}) that
\begin{equation}
e^{-\frac{V_1}{V_0}}e^{-\frac{V_2}{V_0}}\cdot ...\cdot e^{-\frac{V_N}{V_0}}=1/2.
\end{equation}
Since $\sum V_i=V^*$, we find the following scale invariant expression for the dynamics threshold voltage:
\begin{equation}
V^*=\textnormal{ln}(2)V_0 , \label{eq6}
\end{equation}
which is in full agreement with Eq. (\ref{eq13}). Note, however, that Eq. (\ref{eq6}) has been obtained without the assumption
of narrow distribution of memristances.

\subsection{Uniform distribution of threshold voltages}

\begin{figure}[tb]
 \begin{center}
    \includegraphics[width=6.5cm]{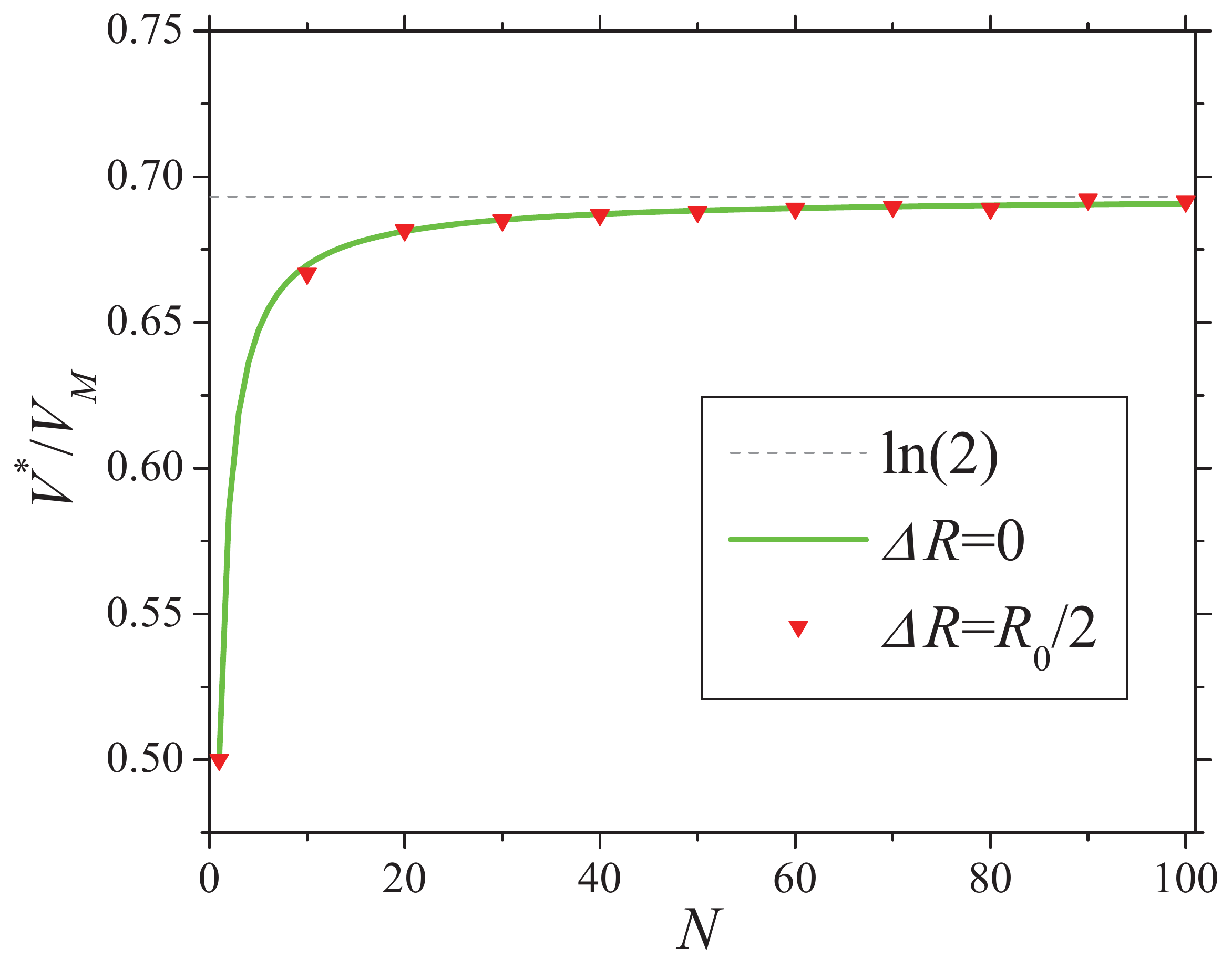}
\caption{\label{fig2} (color online). Dynamics threshold voltage as a function of the number of memristive systems in the network. This plot has been obtained assuming uniform probability distributions of threshold voltages and initial memristances. Specifically, initial memristances are selected randomly in the interval between $R_0-\Delta R$ and $R_0+\Delta R$, and threshold voltages are selected in the interval between $0$ and $V_M$. The solid line was obtained using Eq. (\ref{eq7}).}
\end{center}
\end{figure}

Let us now consider the uniform distribution of the threshold voltages, namely, $f_V(x)=1/V_M$, if $0\leq x\leq V_M$, and
$f_V(x)=0$ otherwise. Here, $V_M$ is the maximum threshold voltage. For the sake of simplicity, we also assume a narrow distribution of initial memristances $R_i=\bar{R}$. Then, from Eq. (\ref{eq5}), it can be found that
\begin{equation}
V^*=NV_M\left( 1-\sqrt[N]{\frac{1}{2}}\right) \xrightarrow{N \rightarrow \infty} \textnormal{ln}(2) V_M. \label{eq7}
\end{equation}
Clearly, the right hand side of Eq. (\ref{eq7}) is again of the form of Eq. (\ref{eq13}). Numerical calculations shown in Fig. \ref{fig2} indicate that a random uniform distribution of initial memristances does not change the result predicted by Eq. (\ref{eq7}).

\subsection{$x\textnormal{exp}(-x)$ distribution of threshold voltages}

\begin{figure}[tb]
 \begin{center}
    \includegraphics[width=6.5cm]{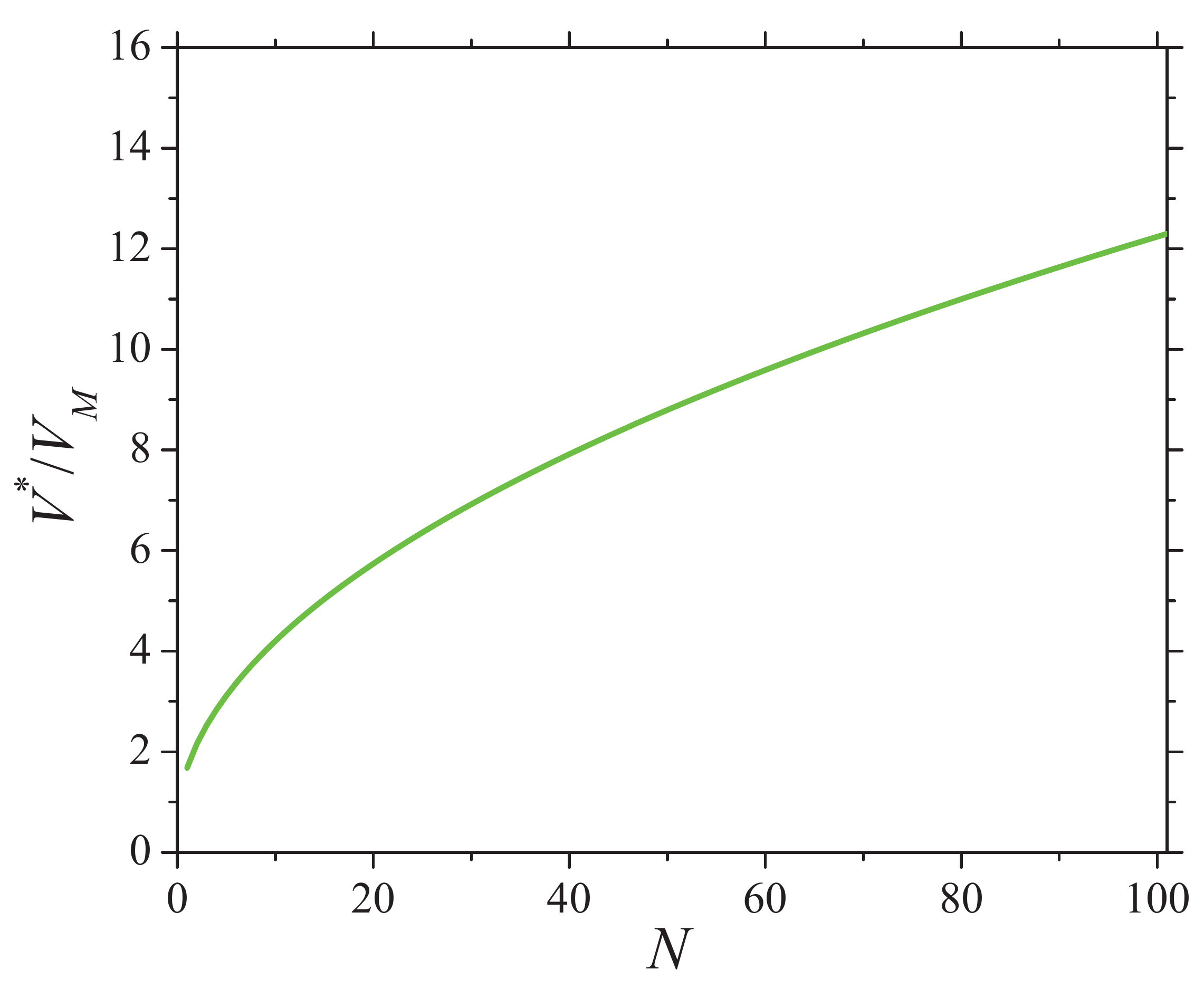}
\caption{\label{fig3} (color online). Dynamics threshold voltage as a function of the number of memristive systems in the network. This plot has been obtained as a solution of Eq. (\ref{eq8}) based on $x\textnormal{exp}(-x)$ distribution of threshold voltages.}
\end{center}
\end{figure}

Finally, we consider an ensemble of memristive systems described by the threshold voltage distribution function $f_V(x)=(x/V_0^2)\textnormal{exp}(-x/V_0)$. The difference with the previously considered cases is that now $f_V(0)=0$ and Eq. (\ref{eq13}) is not applicable.
Assuming a narrow distribution of initial memristances $R_i=\bar{R}$, Eq. (\ref{eq5}) is written as
\begin{equation}
\left(1+\frac{V^*}{NV_0} \right)^Ne^{-\frac{V^*}{V_0}}=1/2. \label{eq8}
\end{equation}
Fig. \ref{fig3} shows a numerical solution of Eq. (\ref{eq8}). The asymptotic
behavior  of the dynamics threshold voltage $V^{*}$ as $N$ tends
to infinity is described by Eq. (\ref{eq12}) with $\gamma=1$, and
$f_0=1/V_{0}^{2}$, i.e.
$V^{*}=V_0 \sqrt{2\ln 2 N}$ as $N\rightarrow\infty$.
Of course, the same result can
be derived  from Eq. (\ref{eq8}) as well.
It is clearly seen that the dynamic threshold voltage $V^*$ does not saturate as in the case of previously considered distributions (see Eqs. (\ref{eq13}), (\ref{eq6}) and (\ref{eq7})). We relate this observation to the absence of memristive systems with zero $V_{tr}$ in the ensemble.

\section{Switching Avalanches in Memristive Ladders} \label{sec5}

Finally, let us consider the possibility of avalanches in memristive networks. By an avalanche, we mean the situation where a single rapid switching event induces switchings in other memristive systems. It follows from our previous considerations (reported in Sec. \ref{sec2}) that avalanches are not possible in purely 1D networks. Indeed, an accelerated switching can not induce an avalanche since the increase in memristance of the switching element reduces the voltages across all other elements in the network. During the slowing-down switching, the inducing switching event is not well characterized as being spread in time.

\begin{figure}[tb]
 \begin{center}
    \includegraphics[width=7.5cm]{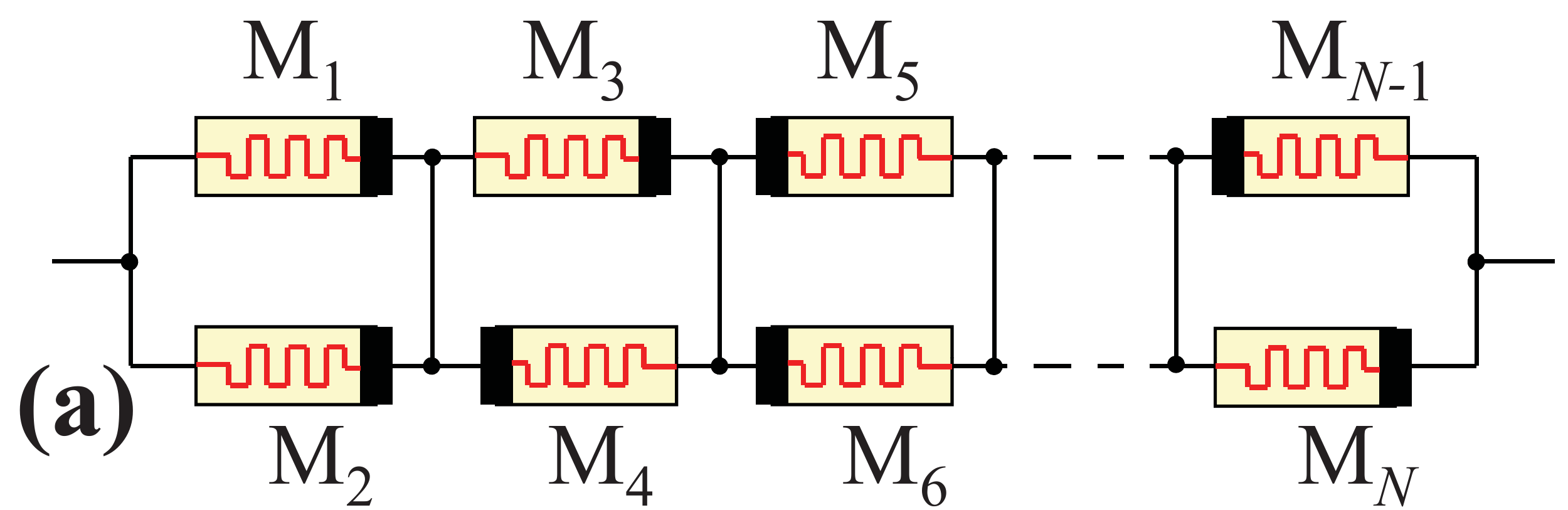}
    \includegraphics[width=8cm]{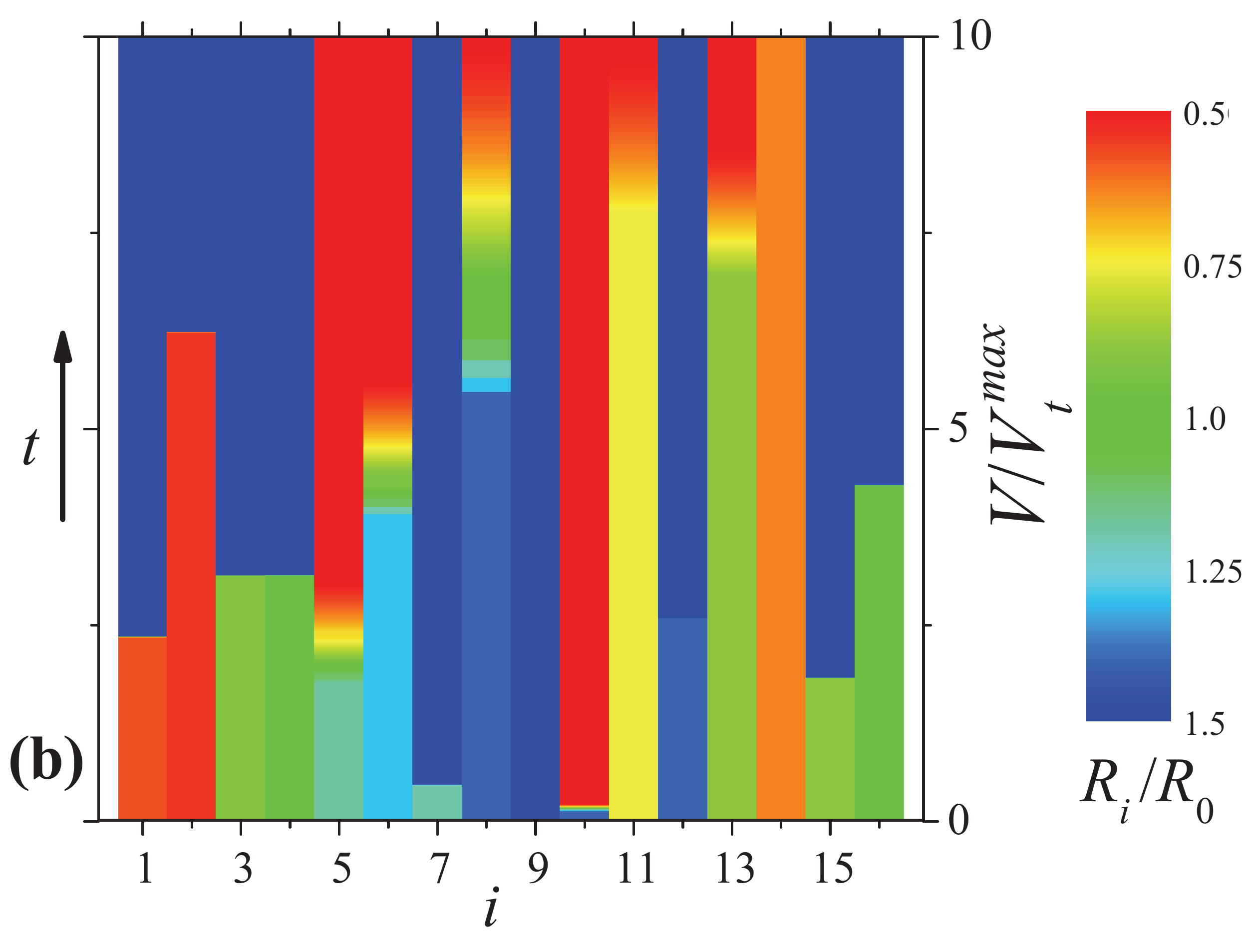}
\caption{\label{fig7} (color online). (a) Memristive ladder consisting of $N$ randomly oriented memristive systems. (b) Individual memristances $R_{i}$ in a memristive ladder of 16 randomly oriented devices as a function of a slowly ramped voltage. The memristance of each device is represented by a color in the band in the vertical direction representing
time evolution. In this simulation, we have used fixed values of $R_{on}=0.5R_0$ and $R_{off}=1.5R_0$, $\beta=2R_0/\left(V_t^{max} \cdot \textnormal{s} \right)$. The threshold voltages and initial memristances of memristive systems have been randomly selected in the interval [0,$V_t^{max}$] and [$R_{on}$,$R_{off}$], respectively.}
\end{center}
\end{figure}

Avalanches, however, are possible in networks of higher dimensions. A memristive ladder (a quasi-one dimensional memristive network) presented in Fig. \ref{fig7}(a) is an example of such situation.
We have performed a numerical simulation of the dynamics of memristive ladder consisting of 2 chains of 8 memristive systems each (16 memristive devices in total) subjected to a slowly ramped voltage.
In the particular realization of memristive devices whose evolution is presented in Fig. \ref{fig7}(b), the simultaneous switching of 3-rd and 4-th memristive systems into the "off" state is an avalanche.

In the memristive ladder, an avalanche is possible in a pair of memristive systems connected in parallel (M$_j$ and M$_{j+1}$, where $j$ is an odd number) that are in the regime of accelerated switching. The switching of the memristive system with a lower $V_t$, say M$_j$, initiates the avalanche (the switching of M$_{j+1}$) if the voltage across these two memristive systems after the switching of M$_j$ exceeds the threshold voltage of M$_{j+1}$. As the switching in the accelerated switching regime occurs fast, effectively, the switchings of  M$_j$ and M$_{j+1}$ are observed at the same value of slowly ramped voltage (see Fig. \ref{fig7}(b) for an example). We expect that the size and role of switching avalanches increase with the increase of network dimensionality. Detailed studies of avalanches in 2D and 3D cases will be reported elsewhere.

\section{Conclusions} \label{sec6}

In conclusion, we have found that one-dimensional memristive networks exhibit several distinctive features that make them quite unlike traditionally studied dynamical networks~\cite{netw_book,motter12a}. The essential difference is related to intrinsic non-local interactions coupled to memory features. Indeed, the order of memristive systems in 1D networks is completely irrelevant since the current through the network is conserved. Two possible types of switching dynamics -- accelerating and decelerating -- have been identified depending on the polarity of memristive elements in the network. We have also demonstrated that the final state of the network depends on the {\it protocol} of how the external voltage is applied, and that switching avalanches can occur in memristive ladders. Additionally, a scale invariance of dynamics threshold voltage has been demonstrated.
Most of our results are universal in the sense that essentially they do not depend on the selected distribution functions (e.g., the acceleration and slowing down of switching, different final states, existence of switching avalanches).
The results reported in this paper are also relevant to networks of memcapacitive and meminductive systems \cite{diventra09a}, while specific features of switching dynamics with such elements may not coincide. In general, we
expect some of these predictions to be shared by arbitrary complex networks with memory.

\section*{Acknowledgements}

This work has been partially supported by NSF grants No. DMR-0802830 and ECCS-1202383, and the Center for Magnetic Recording Research at UCSD.

\bibliography{memcapacitor}
\end{document}